\documentclass[aps, twocolumn, showpacs]{revtex4-1}
\usepackage{amsmath}
\begin{document}
\title{Magnon-phonon conversion experiment and phonon spin}
\author{S. C. Tiwari \\
Department of Physics, Institute of Science, Banaras Hindu University, Varanasi 221005, and \\ Institute of Natural Philosophy \\
Varanasi India\\}
\begin{abstract}
Recent experiment demonstrates magnon to phonon conversion in a YIG film under the application of a non-uniform magnetic field. Light
scattered from phonons is observed to change its polarization state interpreted by the authors signifying phonon spin. In this note
we argue that the experimental data merely shows the exchange of angular momentum $\pm \hbar$ per photon. We suggest that it has physical origin
in the orbital angular momentum of phonons. The distinction between spin and orbital
parts of the total angular momentum, and between phonons and photons with added emphasis on their polarizations is explained. The main 
conclusion of the present note is that phonon spin hypothesis is unphysical.
 
\end{abstract}
\maketitle

\section{\bf Introduction}

Magneto-elastic waves or magnon-phonon excitations have been of interest for various reasons, one of them being the field of spintronics.
A recent experimental study on the magnon-phonon conversion in the ferrimagnetic insulator YIG addresses a question of fundamental importance
whether phonons carry spin \cite{1}. We recall that in 1988 McLellan \cite{2} showed that sharp angular momentum could be attributed to
circular or elliptical phonon polarizations. Note that this angular momentum cannot be identified with the spin of phonon. In this note
we discuss the recent experiment \cite{1} and argue that the measurements show that phonons exchange angular momentum with light but it is
not spin of the phonons. We emphasize that this distinction is not just semantic \cite{3} but of fundamental nature \cite{4}.

The experiment \cite{1} first shows by time-resolved measurements that under the application of a non-uniform magnetic field on a YIG film, spin
wavepackets launched by pulsed microwave signals, convert into elastic wavepackets, i.e. magnon-phonon conversion. Next using wavevector resolved
Brillouin light scattering experiment the measurements show i) magnon-phonon conversion with constant energy and linearly varying momentum, and
ii) the light scattered by the phonons is circularly polarized. The meticulous data presented in Figures (4) and (5) of their paper by the authors
could hardly be doubted. The question that concerns me is regarding their claim, 'that phonons created by the conversion of magnons do carry spin'.

It is true that the change in the polarization state of light involves exchange of angular momentum, for example, to transform linearly polarized
light to circularly polarized light an angular momentum of $\pm \hbar$ per photon is required. However this angular momentum need not be 
associated with the spin of the medium or the light-scattering object. At the macroscopic level, Beth experiment \cite{5} detected a direct
mechanical effect in terms of a torque exerted by a circularly polarized beam of light on a doubly refracting medium which changes the polarization 
state of light. The photon spin angular momentum is transferred to the body of the medium imparting orbital rotation; the aim of the Beth
experiment was, of course, to demonstrate that photons had spin. Holanda et al experiment \cite{1}, on the other hand, assumes photon spin, and
infers that phonons carry spin from microscopic scattering data with photons. The crucial point is that the experiment only proves that angular momentum
in the unit of $\pm \hbar$ is exchanged. It cannot be attributed to the phonon spin: non-zero spin of phonon does not make physical sense.
We argue that the orbital angular momentum of elastic waves or phonons is responsible for angular momentum transfer.

In this note we address the question: Why not spin? First photon physics is briefly reviewed in the next section to highlight the 
intricate relationship between polarization and spin. In section III elementary discussion on phonons shows that photon-phonon analogy
is untenable, and spin cannot be associated with polarized phonons. Further elaboration constitutes section IV. The note ends with a short conclusion.

\section{\bf Photon and light polarization}

A brief review on elementary considerations on the meaning of angular momentum and its decomposition into
orbital and spin parts seems necessary. For a system with rotational symmetry the angular momentum is a constant of motion; if linear momentum
is ${\bf p}$ one defines angular momentum simply as ${\bf r} \times {\bf p}$. In field theory one may construct the expression for angular
momentum from the momentum density of the field or directly calculate the angular momentum density tensor as Noether current from
the rotational invariance of the action. In general, it is useful to separate total angular momentum ${\bf J}$ into orbital and spin components
\begin{equation}
{\bf J} = {\bf L} + {\bf S} 
\end{equation}
For a scalar particle the spin part is zero. For a vector particle, following a textbook discussion \cite{6}, see pp197-198, a simplified picture is 
obtained in terms of ${\bf L}$ that depends on the space or position and spin part that depends on the three components of the vector wavefunction
${\bf V}$ where $3\times 3$ spin matrices $S_x, S_y, S_z$ act on ${\bf V}$, see (27.11) in \cite{6}. Note that similar arguments hold for the second 
quantized field theory. Field ${\bf V}$ becomes an operator
\begin{equation}
{\bf V}_{op} = a_\lambda {\bf V}_\lambda + a_\lambda^\dagger {\bf V}_\lambda^* 
\end{equation}
The annihilation and creation operators $a_\lambda$ and $a_\lambda^\dagger$ for a mode $\lambda$ satisfy the commutation rules
\begin{equation}
[a_\lambda , a_{\lambda^\prime}^\dagger] = \delta_{\lambda \lambda^\prime} 
\end{equation}
The field operators act in Fock space spanned by the Fock state vectors. The orbital angular momentum operator
\begin{equation}
{\bf L} = - i ({\bf r} \times {\bf \nabla})  
\end{equation}
acts on the space-dependence of the mode functions ${\bf V}_\lambda$, whereas the spin operator 
\begin{equation}
{\bf S} = - i \epsilon_{ijk} 
\end{equation}
acts on the components of ${\bf V}_\lambda$.

Photon is a vector particle with rest mass zero and spin one having only two projections - better understood in terms of helicity. Photon
as a quantized electromagnetic radiation field continues to have fundamental questions: gauge-invariance, transversality and Lorentz covariance
are controversial and unsettled issues, see references cited in \cite{4} and also \cite{7, 8}.

Let us try to explain the problem. Classical fields ${\bf E},{\bf B}, A_\mu$ satisfy the wave equation, and one assumes a plane wave
representation. Introducing canonically conjugate field variables canonical quantization is carried out. In the normal mode expansion the annihilation
and creation operators can be defined, and polarization 4-vector $\epsilon_\mu$ comprising of four mutually  orthogonal unit vectors takes
care of the vector nature of the field. In QED a manifest Lorentz covariant quantization results into longitudinal and time-like
photons besides the physical photons.

However in contrast to QED where the electromagnetic potentials $A_\mu$ are fundamental field variables, in quantum optics literature 
the utility of the electric and magnetic field operators is well known. In a simpler field quantization for the radiation field polarization
index $s=1,2$ for transverse fields is sufficient. The normalized eigenstate of the number operator $n_{{\bf k}s} =a^\dagger_{{\bf k}s} a_{{\bf k}s}$
gives the number of photons in the mode $({\bf k}, s)$ as
\begin{equation}
n_{{\bf k}s}| n_{{\bf k}s}> =n_{{\bf k}s} | n_{{\bf k}s}>
\end{equation}
The Fock state is a direct product of number states over all possible modes
\begin{equation}
|\{n\}> = \prod_{{\bf k}s}  | n_{{\bf k}s}>
\end{equation}
The assumption of transverse mode functions, for example, ${\bf A}_\perp$ eliminates longitudinal and time-like photons in quantum optics.

Physical quantities like energy, momentum and angular momentum are obtained using their classical expressions and transforming them to the
quantized field operators. In the classical radiation field theory the Poynting vector ${\bf E} \times {\bf B}$ represents the momentum density
and the total angular momentum density becomes
\begin{equation}
{\bf J} = {\bf r} \times ({\bf E} \times {\bf B}) 
\end{equation}
Separation of (8) into orbital and spin parts can be made similar to (1). The spin angular momentum density
is identified with the expression
\begin{equation}
{\bf S} = {\bf E} \times {\bf A} 
\end{equation}
Regarding spin angular momentum a remarkable result pointed out by van Enk and Nienhuis \cite{9} is worth mentioning. For a circularly polarized
plane wave it is found that the spin operator corresponding to (9) has a simple form
\begin{equation}
{\bf S}^r = \sum_k \frac{\bf k}{|k|} (n_{{\bf k} +} -  n_{{\bf k}-} )
\end{equation}
Here $s=\pm$ for right and left circular polarization. The components of ${\bf S}^r$ commute with each other
\begin{equation}
[S^r_i , S^r_j] =0 
\end{equation}
Authors \cite{9} argue that the spin operator (10) cannot generate polarization rotation of the field, and cannot be 
interpreted as spin angular momentum of photon. Note that Jauch and Rohrlich \cite{10} define Stokes operators satisfying the 
angular momentum commutation rules which provide interpretation of the photon spin \cite{11}.

To conclude this section, in both QED and quantum optics photon spin and the role of polarization state involve intricate issues. One thing is,
however unambiguous, namely that spin angular momentum is an intrinsic property associated purely with the nature of the fields. In fact,
spin for electron also depends only on the Dirac field
\begin{equation}
{\bf \Sigma} = \Psi^\dagger {\bf \gamma} \gamma_5 \Psi
\end{equation}

\section{Phonon spin}

In the abstract of \cite{1} the authors state that, 'while it is well established that photons in circularly polarized light carry
a spin, the spin of phonons has had little attention in the literature'. Now keeping in mind the conceptual problems associated
with photon physics highlighted in the preceding section the photon spin has to be interpreted with great care. The second part of the statement
is, however not correct. The condensed matter literature tacitly accepts phonon to be a zero spin boson, in spite of the transverse modes
and the known polarization of acoustic and optical phonons. Polarization of phonon modes is not related with spin but orbital angular 
momentum \cite{2}. A brief discussion seems useful for the sake of clarity.

Phonons are quantized lattice vibrations; phonon modes are described by wavevector ${\bf k}$, a branch number$j$ and the 
orientation of the coordinate axes \cite{2}. The branch number has two values for crystals with two sub-lattices and there
are two triplets of phonons for acoustic and optical branches. McLellan defines phonon angular momentum in terms of phonon annihilation
and creation operators to be
\begin{equation}
{\bf L}_{ph} = \sum_{\bf{k} j} a_{{\bf k} j} \times a^\dagger_{{\bf k}j} 
\end{equation}
This expression is, as pointed by the author \cite{2}, in agreement with that defined using the displacement vector ${\bf u}_{l\kappa}$
\begin{equation}
{\bf L} =\sum_{l\kappa}   {\bf u}_{l\kappa} \times  {\bf p}_{l\kappa}
\end{equation}
Here the index $l$ corresponds to the unit cell and $\kappa$ for the atom on a sub-lattice. Expression (52) in \cite{2} for
the total angular momentum of the lattice includes that of the rigid body rotation of the crystal.

What are the implications of above discussion? It throws light on the issue of phonon polarization and spin as follows. 

[1] Phonon is a quasi-particle 
having no dynamical field equations like Maxwell field equations for photon. The most crucial point that seems to have gone unnoticed
in the discussions on phonon spin and phonon-photon analogy is that the displacement vector representing lattice vibrations is a real space coordinate.
Canonical quantization and the field operators for phonons are based on the coordinate and momentum, for example those appearing in Eq.(14). On the other
hand, for photon the field variable $A_\mu$ is treated as a coordinate variable, and $\frac{\partial \mathcal{L}}{\partial \dot{A_\mu}}$ is the 
canonically conjugate ``momentum'' variable for the quantization. Here $\mathcal{L}$ is the Lagrangian density of the Maxwell field. 

[2] Phonon polarization is physically entirely different than light or photon polarization. McLellan's analysis clearly establishes the physical significance
of phonon polarization in terms of orbital angular momentum. Isotropic 2D quantum oscillator best illustrates the meaning of polarization of
elastic waves or phonons. In cartesian coordinates the raising and lowering operators separate into 1D oscillators; it is akin to linear
polarization. A circular basis $(a^\dagger_x \pm i  a^\dagger_y)$ formally resembles circular polarization. In the circular basis one gets
well-defined orbital angular momentum of the oscillator.

Transverse modes in paraxial optics also represent physical realization of this example. First order Hermite-Gaussian modes $HG_{10}$ and
$HG_{01}$ are not eigenstates of angular momentum operator (4). However, Laguerre-Gaussian modes
\begin{equation}
LG_0^{\pm 1} =\frac{1}{\sqrt 2} (HG_{10} \pm i HG_{01}) 
\end{equation}
possess sharp angular momentum. Thus phonon polarization is related with orbital angular momentum not spin.

\section{\bf Discussion}
Let us try to elucidate further why photon-phonon analogy is misleading. Photon as a quantized vector field has intrinsic spin one.
Wigner's group theoretical arguments establish that for any massless or light-like particle with non-zero spin there exist only two helicity states.
In the classical picture the intrinsic spin is identified with the vector product of the electric field and the vector potential (9). 
Assumed transverse vector potential leads to the electric field
\begin{equation}
{\bf E}_\perp = - \frac{\partial {\bf A}_\perp}{ \partial t} 
\end{equation}
In the field quantization assuming monochromatic light the electric and magnetic fields are obtained using (16) and ${\bf \nabla} \times {\bf A}_\perp$
respectively, e. g. the expression (6) in \cite{9}.

The oscillations or vibrations around equilibrium position of ions collectively lead to the elastic waves and are analyzed in the 
harmonic approximation in terms of the normal modes. The mode expansion includes wave vector and polarization specifications \cite{12, 13}.
Phonon field is understood in terms of the displacement of a point in the material medium ${\bf u}({\bf r},t)$ and the corresponding momentum
\begin{equation}
{\bf p} = \int \rho \dot{{\bf u}}({\bf r},t) dV
\end{equation}
where $\rho$ is the mass density. Standard coordinate and momentum quantization rule, and plane wave representation yield quantized phonon
field. It is easy to see that expression (14) is just the orbital angular momentum. A deceptive formal analogy with the photon spin expression (9)
is obvious considering expression (16) and using (17) for phonon. Physical interpretation depends on the fundamental distinction between the vector 
potential and the displacement vector since the later is a real space coordinate variable. Thus the suggested interpretation for the phonon
angular momentum corresponding to the circularly polarized modes in \cite{2} seems justified. We remark that in spite of the usage
of phonon polarization in the literature \cite{13}, and transverse polarization in the designation of creation and annihilation operators phonon
spin and vector nature of the phonon field is nowhere mentioned. To avoid confusion, it has to be understood that scalar field could possess
well defined orbital angular momentum and laser light beams with sharp orbital angular momentum have been  extensively studied in the 
literature, see references in \cite{4}. Longitudinal modes have no spin or orbital angular momentum, however linearly polarized light 
could possess orbital angular momentum but not spin. Thus the conventional phonon theory has no analogy with the photon theory, and non-zero phonon 
spin does not make physical sense. The origin of the angular momentum transfer from phonon to photon in the reported experiment \cite{1}
may be logically attributed to the orbital angular momentum of phonons.

In a hypothetical scenario assuming phonon has spin one it would be of interest to find its physical consequences. I think electron-phonon
interaction and Cooper pair formation via phonon mediated electron-electron interaction  may be re-examined: phonon creation and annihilation
operators \cite{13} could be generalized for the circularly polarized modes in the interaction Hamiltonian and treated as spin one particles. 
There is another problem in superconductivity highlighted by Post \cite{14}, namely the angular momentum conservation in a superconducting ring.
Though Post sets the problem in the form of Onsager-Feynman controversy he offers insightful discussion on the mechanism of the angular
momentum balance when supercurrent in a ring vanishes as the temperature is raised above the transition temperature. Note that
Post rules out any role of lattice, therefore it may be of interest to examine the role of phonon spin in this problem.

We could, of course, explore new physics or unconventional ideas \cite{4}. Departing from the phonon picture new kind of field excitations in
the spirit of Cosserat medium was suggested in \cite{4}. Analogy of displacement vector with the vector potential is not justified, however
the velocity field in a rotating fluid may be treated as a vector potential: postulating rotating space-time fluid with nontrivial topology of 
vortices we have re-interpreted the electromagnetic field tensor as the angular momentum (density or more appropriately flux) of 
photon fluid \cite{8}, and proposed a topological photon \cite{15}. Note that the net angular momentum of the microscopic particles 
in the rotating fluid implies antisymmetric stress tensor. Such speculations relate spin with topological invariants.

\section{Conclusion}

It has been pointed out \cite{4} that phonon angular momentum discussed in \cite{16} is ambiguous as compared to that discussed in \cite{2}.
We have shown that non-zero phonon spin hypothesis and phonon-photon analogy \cite{17} are conceptually flawed, giving further support to the arguments 
presented in \cite{4}. Phonon spin has no experimental evidence.
The correct physical interpretation of the reported experiment \cite{1} is that orbital angular momentum of phonons is exchanged with light beam resulting 
into the change in the polarization of the light.

{\bf Acknowledgments}

I thank S. Streub for raising specific questions on the photon-phonon analogy. I also acknowledge correspondence with S. M. Rezende, M. Wakamatsu,
and A. Hoffmann, and conversation with D. Sa and V. S. Subrahmanyam.

\end{document}